\documentclass[fleqn,usenatbib]{mnras}

\usepackage{newtxtext,newtxmath}

\usepackage[T1]{fontenc}
\usepackage{ae,aecompl}

\usepackage{graphicx}	
\usepackage{amsmath}	
\usepackage{amssymb}	

\title[$B$-$n$ relation]{The Density-Magnetic Field Relation in the Atomic ISM}

\author[Gazol \& Villagran]{
A. Gazol,$^{1}$\thanks{E-mail: a.gazol@irya.unam.mx}
M. A. Villagran,$^{1}$
\\
$^{1}$Instituto de Radioastronom{\'\i}a y Astrof{\'\i}sica, UNAM
}

\date{Accepted XXX. Received YYY; in original form ZZZ}

\pubyear{2018}

\begin{document}
\label{firstpage}
\pagerange{\pageref{firstpage}--\pageref{lastpage}}
\maketitle

\begin{abstract}
We present numerical experiments aimed to study the correlation between the magnetic field strength, $B$, and the density,  $n$, in the cold atomic interstellar medium (CNM). We analyze 24 magneto-hydrodynamic models with different initial magnetic field intensities ($B_0=$0.4, 2.1, 4.2, and 8.3~$\mu$G) and/or mean densities (2, 3, and 4~cm$^{-3}$), in the presence of driven and decaying turbulence, with and without self-gravity, in a cubic computational domain with 100~pc by side. Our main findings are: i) For forced simulations, which reproduce the main observed physical conditions of the CNM in the Solar neighborhood, a positive correlation between $B$ and $n$ develops for all the $B_0$ values. ii) The density at which this correlation becomes significant ($\lesssim 30$~cm$^{-3}$) depends on $B_0$ but is not sensitive to the presence of self-gravity. iii) The effect of  self-gravity, when noticeable, consists of producing a shallower correlation at high densities, suggesting that, in the  studied regime, self-gravity induces motions along the field lines. iv) Self-gravitating decaying models where the CNM is subsonic and sub-Alfv\'enic with $\beta\lesssim 1$ develop a high density positive correlation whose slopes are consistent with a constant $\beta(n)$. v) Decaying models where the low density CNM is subsonic and sub-Alfv\'enic with $\beta>1$ show a negative correlation at intermediate densities, followed by a high density positive correlation.
\end{abstract}




\section{Introduction}
The relation between the hydrogen particle density and the magnetic field intensity in the interstellar medium has been  studied during more than three decades through Zeeman effect measurements from different tracers corresponding to different density ranges. For the atomic interstellar gas, and more specifically for the cold neutral medium (CNM) traced through the 21cm HI line, a weak or null correlation has been reported by \cite{HeilesTroland86}. For molecular gas, in contrast, a positive correlation has been found from OH and CN data \citep{Crutcher99}. The emergence of a regime where the magnetic field is correlated  with density has been traditionally interpreted as the result of self-gravity effects becoming important in generating gas motions perpendicular to the field lines, which in turn produce a compression of the freezing-in magnetic field. However alternative explanations, based exclusively on the interactions between gas motions and magnetic field fluctuations, have also been proposed. On one hand, scaling relations between the density and the magnetic pressure produced by different magnetohydrodynamic (MHD) wave modes  in the non-linear regime have been analytically found \citep{McKeeZweibel95,PassotVS03}. On the other hand, numerical simulations of compressible magnetized turbulent isothermal gas have shown that a positive  correlation between the density and the magnetic field intensity develops when supersonic motions are present, both in the super-Alfv\'enic as well as in the sub-Alfv\'enic case \citep{PadoanNordlund99,Ostrikeretal01,Burkhartetal09,Collinsetal11}.    

The most recent observational version of this relation has been given by \cite{Crutcheretal10}, who compiled data from the available surveys tracing gas with densities going from few tens to 10$^6$~cm$^{-3}$. They used these data  to find the most probable maximum value for the total magnetic field intensity in the case where no correlation is assumed for densities below a transition density $n_{Ht}$, and a positive correlation, with an exponent $\alpha$, is assumed for higher densities. In that work, Bayesian statistics is used to find $n_{Ht}=300$~cm$^{-3}$, $\alpha=0.65$, and the maximum magnetic field value for the no correlation regime, $B_{max}=10$~$\mu$G. The value of $n_{Ht}$ corresponds to the mean hydrogen particle density in giant molecular clouds, favoring the previously mentioned interpretation in terms of the effects of self-gravity. Nevertheless this value lies also in the density range corresponding to the CNM gas, where supersonic motions are thought to be common, cooling is non-negligible, and self-gravity is usually sub-dominant.  

Additionally, recent observations have found a preference for the projected magnetic field on the plane of the sky to be parallel to matter structures in the CNM, traced by HI \citep{Clarketal14} as well as by dust emission \citep{PlanckXXXII}. This alignment could be a consequence of the presence of super-Alfv\'enic turbulence, a regime in which gas compressions produced by supersonic motions are able to generate an enhancement of the magnetic field component perpendicular to motions \citep{PlanckXXXII}, suggesting the possible presence of a correlation between $B$ and $n_H$ in the CNM. The same trend in the alignment between the plane of the sky magnetic field and the dense structures has been recovered in numerical simulations of different types: colliding flows in a magnetized medium \citep{InoueInutsuka16}, and  driven turbulence models \citep{Marco17}. In the latter work we have found that the alignment does also hold for the three-dimensional local magnetic field, reinforcing the possibility of magnetic field compression. This brings up the question of how the magnetic field intensity is related with density in a gas with thermal and dynamical properties similar to those reported for the CNM and where regions are predominantly magnetically sub-critical \citep{HeilesTroland2004}. 

Zeeman observations from HI absorption are consistent with an uncorrelated magnetic field intensity with density, but they are not conclusive. Actually, the available data, analyzed in \cite{Crutcheretal10}, make it hard to distinguish between the following three different cases: i) all the observed clouds have the same total magnetic field, ii) the total magnetic field intensity of each region is drawn from a uniform distribution, or iii) the distribution of total magnetic field intensity leading to the observed line-of-sight field is the one obtained with the numerical model presented by \cite{Falcetaetal2008}. This fact, along with the large uncertainties on the line of sight magnetic field measures and on the volumetric density determination --from the spin temperature of each component and a uniform pressure (3000~Kcm$^{-3}$) assumption--, and with the alignment results previously mentioned, constitute a strong motivation for a detailed investigation of the $B$-$n_H$ relation in physical conditions similar to those of the CNM.

This relation has also been examined in the context of non isothermal simulations. The simulations utilized to this end can be roughly divided in two groups; on one hand models considering a supernovae driven ISM \citep{BalsaraKim2005,deAvillez2005,Hilletal12,IffrigHenn14,Pardi2017} and, on the other, models aimed to investigate molecular cloud formation \citep{Hennebelleetal2008,Heitschetal09,Banerjee09,InoueInut2012,Fogerty2016,Wuetal17}. In the first group, numerical requirements needed to include all the relevant physics and/or to take into account large scales ($\geqslant 1$~kpc) do not allow to use enough resolution to study the CNM gas. In the second group, the focus has been put on the study of the properties and the evolution of molecular gas, which needs to resolve small scales in high density regions. For this reason, these studies frequently  use numerical schemes that give resolution priority to dense regions (i.e. Smoothed Particle Hydrodynamics or Adaptive Mesh Refinement) but treat more diffuse CNM-like gas with relatively low resolution. In the more recent works of this group, where the diffuse gas has enough resolution, no systematic attention has been payed to the $B$-$n$ relation in this gas.  

The aim of the present work is thus to make a systematic numerical study of the relationship between the density and the magnetic field intensity in the CNM, by analyzing models where the relevant physical ingredients of this kind of gas in the Solar neighborhood are taken into account, and where an adequate resolution is used. 

The plan of the paper goes as follows. In the next section we briefly summarize some previous theoretical results concerning the relation between $B$ and $n$, then, in section \ref{sec:sim} we describe the set-up of the models that we analyze in section \ref{sec:res}. These results are discussed in section \ref{sec:disc} and, finally,  a conclusion is presented in section \ref{sec:conc}.

\section{Theoretical predictions}\label{sec:betas}
In this section we briefly summarize some theoretical predictions for the $B-n$ relation that are useful for the discussion and interpretation of our results.  

The case of an isothermal spherical cloud dominated by gravity and contracting under its effects has been studied by \cite{Mestel65}, who predicted $B\propto n^{2/3}$. The same behavior is expected whenever isotropic gravitational contraction is present in flux-freezing conditions, i.e. when the magnetic field is not strong enough to modify the geometry of the contracting region. When contraction happens under a constant ratio of thermal ($P_{th}$) to magnetic ($P_B$) pressure, $\beta=P_{th}/P_{B}$, and $P_{th}\propto n^{\gamma}$, then $B\propto n^{\gamma/2}$ \citep{Mousco91}. Thus, for the particular case of isothermal gas at constant $\beta$, $B\propto n^{1/2}$. In this case the mass to magnetic flux ratio of
the region increases as the contraction proceeds because this process is not isotropic. This behavior is expected when the presence of the magnetic field has some influence on the contraction geometry.

When gravitational contraction is present, the magnetic field can also be amplified by the small-scale turbulent dynamo produced by the conversion of potential energy into  turbulent kinetic energy \citep[e.g.][]{Federrathetal2011,Federrath16}. As this process is highly dependent on the magnetic diffusivity, its contribution to numerical results is highly dependent on the resolution. \cite{Federrathetal2011} argued that a minimum number of 30 grid cells per Jeans length is required in order to observe some effects of this kind of dynamo.

The transition from a no correlated regime at low densities to a regime characterized by a correlation with a value of $2/3$ for the exponent has also been interpreted as the result of diffusion caused by turbulent magnetic reconnection \citep{Lazarianetal2012}. In that work the authors present a simple reconnection diffusion model to show that this process can effectively remove magnetic field in molecular cloud envelopes and argue that, being more efficient at low densities, there is a scale range where reconnection diffusion can act on timescales shorter than those of gravitational collapse. 

On the other hand, the relationship between the magnetic pressure produced by undamped Alfv\'en  waves and the density has been studied by \cite{McKeeZweibel95}. They found that, in hydrostatic equilibrium, this pressure, $P_A$, is proportional to $n^{1/2}$; while in a slow varying gas (moderate Alfv\'enic Mach number, $M_A$, and long wavelength), $P_A\propto n^{3/2}$ for three cases: i) isotropic Alfv\'en waves, ii) spatially uniform density, and iii) self-similar contraction. Finally they argue that in the presence of a shock, the exponent relating $P_A$ with $n$ could be as large as 2, depending on the shock direction. These results have been recovered by \cite{PassotVS03}, who performed a perturbation analysis on top of circularly polarized Alfv\'en waves. They also investigated the scaling of the magnetic pressure produced by magneto-sonic modes  with density. To this end they used simple waves \citep{Bolliat70,webbetal96} in a slab geometry, and found that the magnetic pressure associated with the fast mode is $P_B\propto n^2$, while for the slow mode $P_B\propto c_1-c_2n$ (with $c_1$ and $c_2$ being constants). The latter expression implies that, when slow mode is dominant, $P_B$ is independent of density at low enough densities and it decreases $n$ for higher values.  Comparing the different scalings resulting from their analysis \cite{PassotVS03} concluded that at low densities the $B$-$n$ relation is dominated by the slow mode and no correlation is expected, while at high enough densities, when the contribution of the slow mode disappears, a positive correlation develops due to the domination of the Alfv\'en and the fast modes. Note however that the application of this analysis to a gas with physical properties similar to those of the CNM remains unclear. Mainly because magneto-sonic scalings are derived in a limit where important density fluctuations with weak magnetic field perturbation, a situation expected in the presence of supersonic motions and a strong magnetic field, can not be considered. Note that a negative correlation has been numerically observed for three-dimensional isothermal subsonic and sub-Alfv\'enic turbulence \citep{Burkhartetal09}, although its presence seems to depend on the  driving scheme, and more specifically on the correlation time associated with successive forcing events, whose effects become relevant at transonic velocities, \cite{Yoonetal2016}. 

\section{The simulations}\label{sec:sim}
In this paper we analyze the results of 24 simulations which correspond to 6 groups with 4 different configurations: with/without self-gravity and forced/decaying turbulence. Each group has different initial magnetic field intensity, $B_0$, and/or different initial density, $n_0$ (see table \ref{tab:runs}). Forced simulations are part of models analyzed in \cite{Marco17} (we keep the same labels for these models), which have been restarted with turbulent forcing turned off in order to produce the decaying turbulence models that we present in this work (labeled with an additional D). The main physical properties of the CNM in the Solar neighborhood are reproduced by forced models, particularly by those with $n_0=2$~cm$^{-3}$ and $B_0=4.2$ and 8.3~$\mu$G. In fact, as discussed in detail in \cite{Marco17}, the Alfv\'enic Mach number histogram for the CNM peaks at 1.7 and 1, for 4.2 and 8.3~$\mu$G, respectively. These values encompass the observationally determined estimation of 1.4 for a mean magnetic field intensity of 6~$\mu$G \citep{HeilesTroland2003}. The corresponding sonic Mach number histograms for the CNM resulting from all the forced models have peaks at moderately supersonic values ($\approx 3.6$), which are also in agreement with the observational estimation by \cite{HeilesTroland2003}.

All the simulations result from the integration of the ideal MHD equations in the presence of cooling, using 512$^3$ grid cells to represent a periodic cubic box of 100~pc by side, filled with gas at mean density and mean pressure laying in the thermally unstable regime \citep{Field65} of the atomic interstellar gas. For this, we use a TVD code based on the one presented by \cite{kimetal99}. In the forced models, the initial uniform density, $n_0$, and temperature, $T_0$, have thermal equilibrium values according to our cooling function. The cooling function is a fit, based on \cite{Wolfireetal03} results for the solar galactocentric radius \citep{GazolVillagran16}, which assumes a constant heating rate  $\Gamma=22.4\times10^{-27}$~erg~cm$^3$s$^{-1}$.
Driven models start at rest with a uniform magnetic field, $\mathbf{B_0}=B_0\mathbf{x}$, and are forced in the Fourier space with solenoidal forcing applied at a fixed wavenumber corresponding to a physical scale of 50~pc. The amplitude of velocity perturbations is fixed by a constant injection rate of kinetic energy, which we chose so as to approximately get the same rms velocity for all the driven models. 
The self-gravitating models are very gravitationally stable, having Jeans lengths of 552, 382 and 278~pc, for $n_0=2$, 3, and 4~cm$^{-3}$, respectively. For the resolution we use, the Truelove criterion \citep{Truelove1997} to avoid artificial fragmentation is satisfied for densities below $\approx 6350$~cm$^{-3}$, while the maximum density to resolve small-scale dynamo produced by gravitational induced motions \citep{Federrathetal2011} is  $\approx 5000$~cm$^{-3}$.
For further details concerning the model and the code see \cite{Marco17}, \cite{GazolKim2010}, and \cite{kimetal99}. 

\begin{table}
 \caption{Run parameters}
 \label{tab:runs}
 \begin{tabular}{llllcc}
  \hline
  Model&$n_0$ & $T_0$  & $B_0$  & Forcing & Self-gravity\\
  & cm$^{-3}$ & K & $\mu$G & &\\
  \hline
  B01S&    2& 1500& 0.4 & on  & off \\
  B01G&    2& 1500& 0.4 & on  & on  \\
  B01SD&   2& 1500& 0.4 & off & off \\
  B01GD&   2& 1500& 0.4 & off & on  \\
  B05S&    2& 1500& 2.1 & on  & off \\
  B05G&    2& 1500& 2.1 & on  & on  \\
  B05SD&   2& 1500& 2.1 & off & off \\
  B05GD&   2& 1500& 2.1 & off & on  \\
  B10S&    2& 1500& 4.2 & on  & off \\
  B10G&    2& 1500& 4.2 & on  & on  \\
  B10SD&   2& 1500& 4.2 & off & off \\
  B10GD&   2& 1500& 4.2 & off & on  \\
  B20S&    2& 1500& 8.3 & on  & off \\
  B20G&    2& 1500& 8.3 & on  & on  \\
  B20SD&   2& 1500& 8.3 & off & off \\
  B20GD&   2& 1500& 8.3 & off & on  \\
  B10Sn3&  3& 733 & 4.2 & on  & off \\
  B10Gn3&  3& 733 & 4.2 & on  & on  \\
  B10SDn3& 3& 733 & 4.2 & off & off \\
  B10GDn3& 3& 733 & 4.2 & off & on  \\
  B10Sn4&  4& 518 & 4.2 & on  & off \\
  B10Gn4&  4& 518 & 4.2 & on  & on  \\
  B10SDn4& 4& 518 & 4.2 & off & off \\
  B10GDn4& 4& 518 & 4.2 & off & on  \\
  \hline
 \end{tabular}
\end{table}
\section{Results}\label{sec:res}
In this section we analyze the relation between the magnetic field intensity and the density for data resulting from the models described in the previous section. As a first approach, we take into account all the gas in the box and we analyze the $B$-$n$ relation at densities comparable with those of the CNM, making no distinction between individual dense structures. In order to  characterize the behavior of $B$, we use the median magnetic field at each density value, $B_m$, computed from the two-dimensional (2d) histogram. Note that we have tested the effects of using the mean value of $B$ instead of its median value and we did not find important differences for the results presented below. For the sonic ($M_s$) and  the Alfv\'enic Mach numbers we also utilize their median value at each density resulting from the corresponding 2d histogram.

\subsection{The effect of varying $B_0$}
\subsubsection{Forced Turbulence}\label{sec:resMain}
We first present results from forced models with $n_0=2$~cm$^{-3}$ and different initial magnetic field intensities with and  without self-gravity. 

\begin{figure}
 \includegraphics[width=\columnwidth]{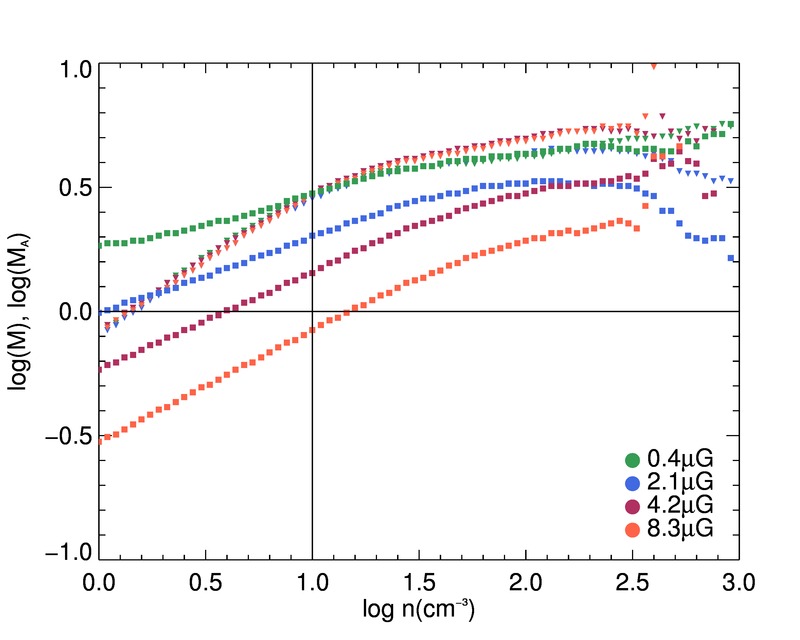}
 \caption{Sonic (triangles) and Alfv\'enic (squares) Mach numbers for forced models without self-gravity.
 The symbols represent the median value of $M_s$ and $M_A$ at each density. Black straight lines at  $\log n=1$ and $\log M=0$ are included as references.}
 \label{fig:Mmain}
\end{figure}

\begin{figure}
 \includegraphics[width=\columnwidth]{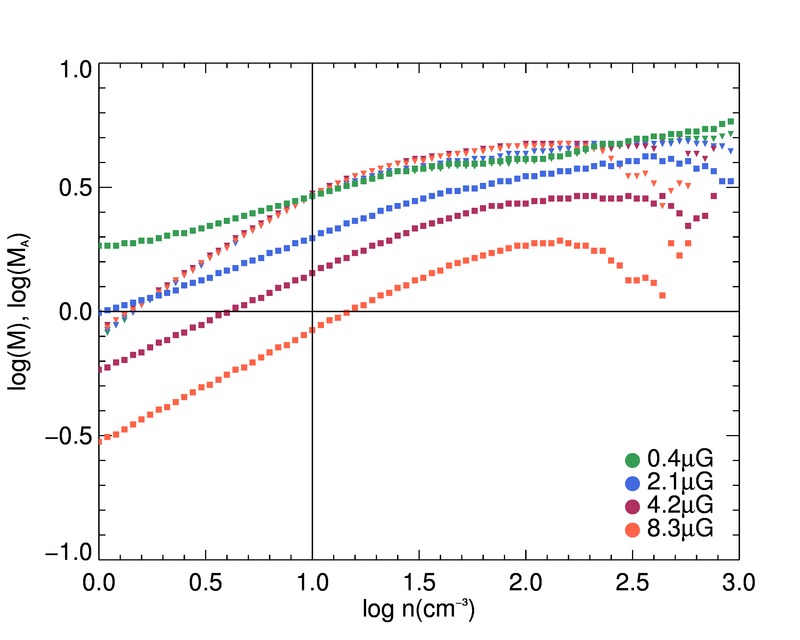}
 \caption{Sonic (triangles) and Alfv\'enic (squares) Mach numbers for forced models with self-gravity. The symbols represent the median value of $M_s$ and $M_A$ at each density. Black straight lines at  $\log n=1$ and $\log M=0$ are included as references.}
 \label{fig:MmainG}
\end{figure}

From figures \ref{fig:Mmain} and \ref{fig:MmainG}, where the median value of $M_s$ and $M_A$ as a function of density is displayed for non self-gravitating and self-gravitating models, respectively, it can be seen that for all the models in this set, the gas with densities similar to those of the CNM ($n\gtrsim 10$~cm$^{-3}$) is moderately supersonic and  super-Alfv\'enic, in agreement with observations \citep{HeilesTroland2003,HeilesTroland2004}. As the $\beta$ plasma parameter is related with $M_s$ and $M_A$ by $\beta=2M_A^2/M_s^2$, the same figures imply that in our models, the dense gas has $\beta\lesssim 1$ which is also consistent with observations \citep{HeilesCrutcher2005}.  

\begin{figure}
 \includegraphics[width=\columnwidth]{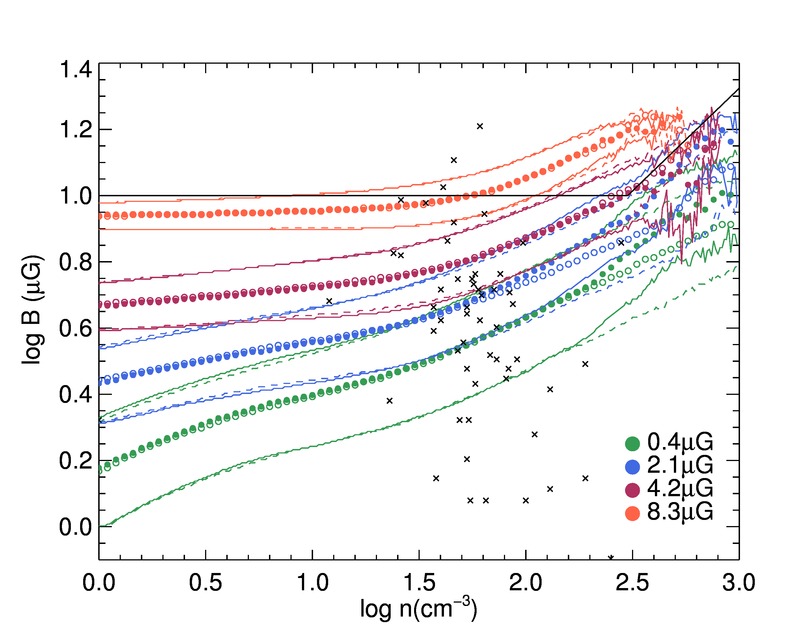}
 \caption{$B$-$n$ relation for forced models with $n_0=2$~cm$^{-3}$ as traced by the median of the 2d histogram. Filled(empty) circles are for models without(with) self-gravity. For each initial magnetic field continuous(dashed) lines are for $B$ values containing 25$\%$ of points below and above the median for models without(with) self-gravity. Black lines are the most probable maximum values found by \protect\citet{Crutcheretal10}, while black crosses are the HI data from \protect\cite{HeilesTroland2004}.}
 \label{fig:Bromain}
\end{figure}

In figure \ref{fig:Bromain} we display $B_m$ as a function of density for models  with (empty circles) and  without (filled circles) self-gravity for the four $B_0$ values that we consider.  These plots have been obtained from averaging over the last 2.9 turbulent crossing times. First of all, it can be noticed that this set of models covers a region in the $n-B$ plane in which a large number of the available HI Zeeman data \citep[black crosses]{HeilesTroland2004} is included. The second noticeable feature is that the magnetic field intensity grows with density for $\log n\geqslant 1.5$ for all the $B_0$ values we consider. For low $B_0$ models, the rise in $B_m$ starts at smaller densities, where the gas motions are already predominantly super-Alfv\'enic (see figures \ref{fig:Mmain} and \ref{fig:MmainG}). As $B_0$ increases, the density at which $B_m$ starts to grow with density, as well as the density at which gas motions become mainly super-Alfv\'enic, shift to larger values. Finally, figure \ref{fig:Bromain} shows that the effect of self-gravity is not appreciable for high $B_0$ models but, as expected, it becomes stronger as $B_0$ decreases. However, in this case the increment of $B_m$ with $n$ is shallower when self-gravity is included, which is an unexpected behavior that suggests the presence of density enhancements without the corresponding increase of $B$. This could be a consequence of gravitational force producing motions parallel to the field lines, i. e. motions that do not compress the magnetic field lines. Unfortunately our simulations do not have enough high density points in order to quantify this effect. Although the growth of $\log B_m$ with $\log n$ does not seem to follow a linear relation, for reference, we computed the slope of a linear least-square fit for $2\leq\log n\leq 2.6$. For this density range, the mean polytropic index according to our thermal equilibrium curve is 0.7. For the non self-gravitating runs we find a slope $\alpha$ of 0.43, 0.38, 0.28, and 0.27 for $B_0=0.4$, 2.1, 4.2, and 8.3~$\mu$G, respectively; while for the corresponding self-gravitating models $\alpha=0.30$, 0.28, 0.32, and 0.32. These values indicate that, for the low $B_0$ cases, the presence of self-gravity makes $\alpha$ going from values $>\gamma/2$, for which the magnetic pressure increases faster with density than thermal pressure (see section \ref{sec:betas}), to values $<\gamma/2$, where the opposite is true, i. e. the presence of self-gravity causes a qualitative change in the $B-n$ relation for low $B_0$ values. 
\subsubsection{Decaying Turbulence}\label{sec:resMainNF}
\begin{figure}
 \includegraphics[width=\columnwidth]{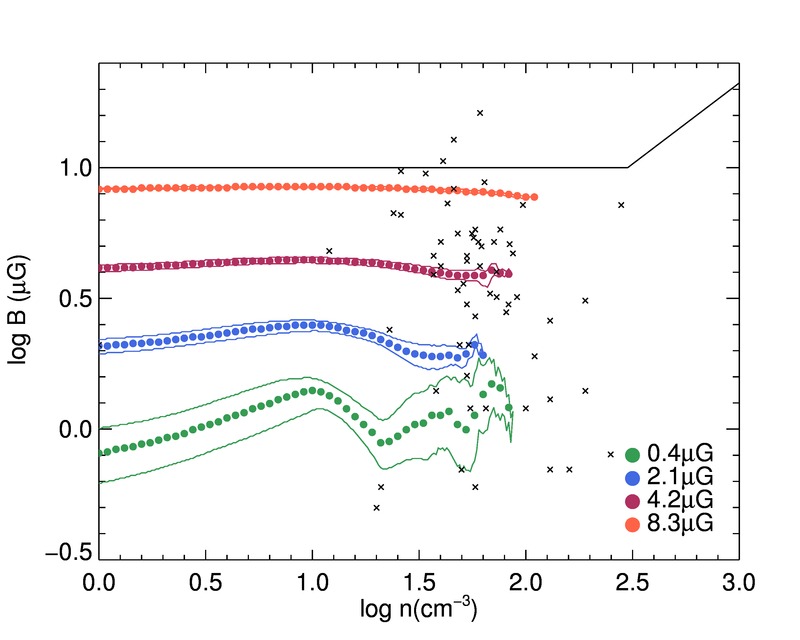}
 \caption{$B$-$n_H$ relation for decaying models with $n_0=2$~cm$^{-3}$ without self-gravity. For each initial magnetic field continuous(dashed) lines are for $B$ values containing 25$\%$ of points below and above the median for models without(with) self-gravity. Black lines are the most probable maximum values found by \protect\citet{Crutcheretal10}, while black crosses are the HI data from \protect\cite{HeilesTroland2004}.}
 \label{fig:BromainNF}
\end{figure}

\begin{figure}
 \includegraphics[width=\columnwidth]{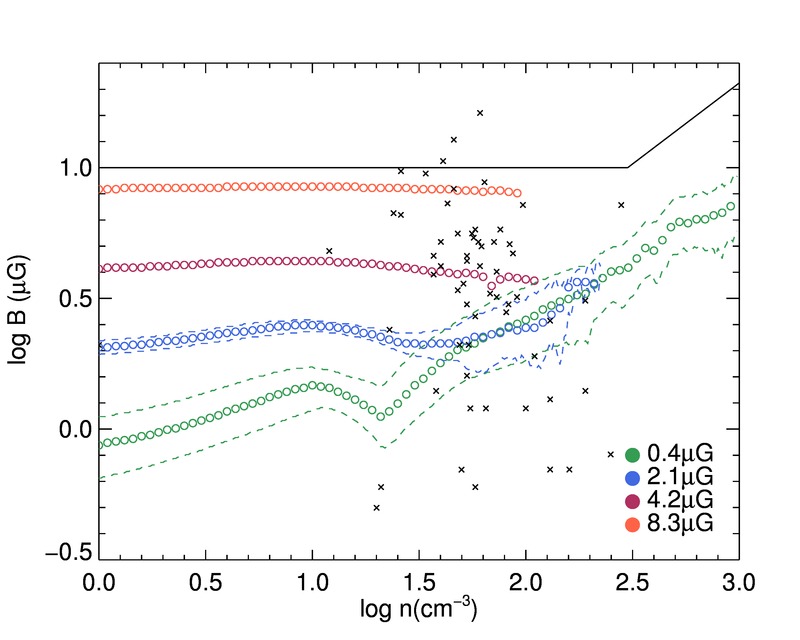}
 \caption{The same as figure \ref{fig:BromainNF} for decaying self-gravitating models with $n_0=2$~cm$^{-3}$.}
 \label{fig:BromainNFG}
\end{figure}

\begin{figure}
 \includegraphics[width=\columnwidth]{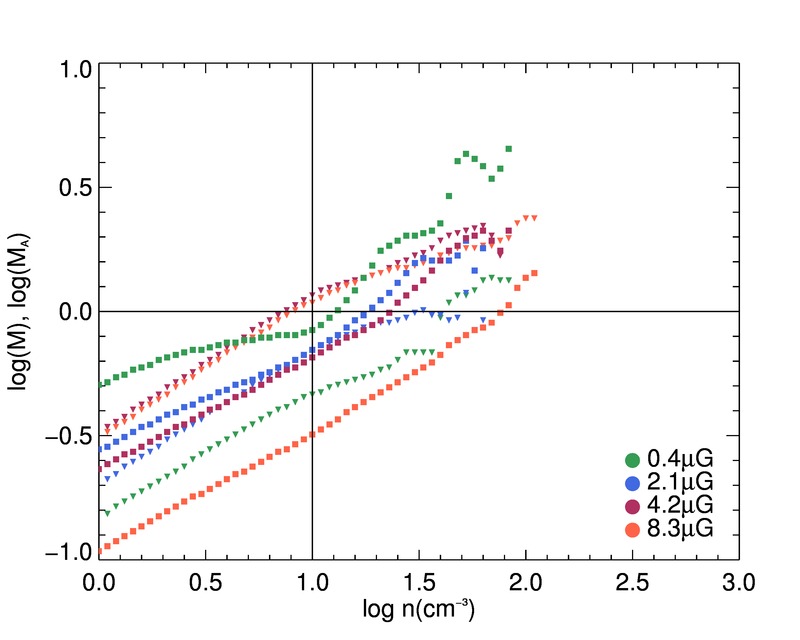}
 \caption{Sonic (triangles) and Alfv\'enic (squares) Mach numbers for decaying models with $n_0=2$~cm$^{-3}$ and without self-gravity.}
 \label{fig:MmainNF}
\end{figure}

\begin{figure}
 \includegraphics[width=\columnwidth]{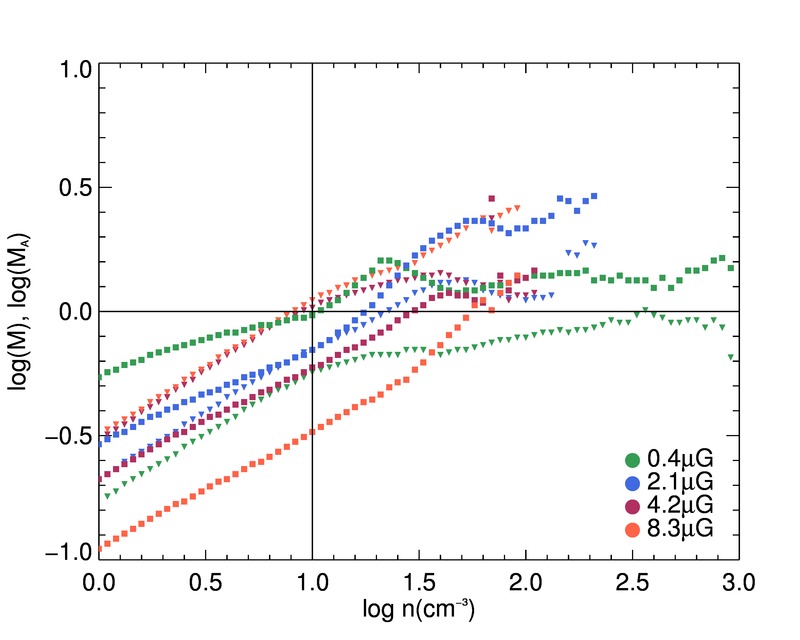}
 \caption{Sonic (triangles) and Alfv\'enic (squares) Mach numbers for self-gravitating decaying models with $n_0=2$~cm$^{-3}$.}
 \label{fig:MmainNFG}
\end{figure}

Although the models discussed in the previous section satisfactorily reproduce a number of observed CNM properties \citep[see also][]{Marco17}, it seems natural to raise the questions of how $B$ is related to $n$ in the case of decaying turbulence, and which are the effects of self-gravity in this case. The $B$-$n$ relation, as measured again by the median value of the two-dimensional histogram, for unforced restarts of models presented in section \ref{sec:resMain} are shown in figures \ref{fig:BromainNF} and \ref{fig:BromainNFG} for the non self-gravitating and the self-gravitating group, respectively; while the corresponding plots for $M_s$ and $M_A$ as a function of $n$ are displayed in figures \ref{fig:MmainNF} and \ref{fig:MmainNFG}. These figures have been done for times at which more than 85 per cent of the initial kinetic energy $E_0$ has been dissipated. In order to improve the statistics we average over ten snapshots during which the fraction of dissipated energy varies among runs between $0.2$ and $2$ per cent of $E_0$.  It can be seen that the models with a relatively high magnetic field, for which the CNM remains supersonic but becomes sub-Alfv\'enic for the low density zone, show an almost flat $B_m$ with small ($\lesssim 100$~cm$^{-3}$) maximum densities, and their behavior does not seem to be affected by the presence of self-gravity. In contrast, in simulations with low $B_0$, for which the CNM is predominantly subsonic and sub-Alfv\'enic at low densities, $B_m$  varies with $n$ in a more complex way which is affected by self-gravity. In fact, at densities below the CNM minimum density these two models show a slight ($B_0=2.1$~$\mu$G) or moderate ($B_0=0.4$~$\mu$G) increase of $B_m$ with $n$, but in a density range starting at $n\approx 10$~cm$^{-3}$ $B_m$ clearly decreases with $n$. The width  of this range is smaller for the lowest $B_0$ value and seems to be independent of self-gravity. In both cases, with and without self-gravity, $B_m$ starts to grow with $n$ again at large enough densities. Non self-gravitating low $B_0$ models reach maximum density values even lower than those attained by models with a larger initial magnetic field. In contrast, self-gravitating low $B_0$ models show a sustained growth of the magnetic field intensity with density over a density range that, for $B_0=0.4$~$\mu$G (green symbols in figure \ref{fig:BromainNFG}) goes up to $n\approx 10^3$~cm$^{-3}$. From figure \ref{fig:MmainNFG} it can be seen that in this case, for densities at which $B_m$ increases with $n$, the median value of the sonic Mach number slowly increases with $n$ while $M_A$ remains relatively constant at trans-Alfv\'enic values, implying a decline in the $\beta$ value. For this model, the slope of a linear fit for the $B$-$n$ relation at $10^2\leq n \leq 10^3$~cm$^{-3}$ is $\approx 0.41$, which is larger than $\gamma/2$, consistent with a magnetic pressure growing faster with $n$ than thermal pressure. 
\subsection{The effect of varying the initial density}\label{sec:HD}
In this section we present results from the models with higher mean density and $B_0=4.2$~$\mu$G both in the forced and in the decaying case. These models are useful to further study the effects of self-gravity. 
\subsubsection{Forced Turbulence}\label{sec:resHD}
\begin{figure}
 \includegraphics[width=\columnwidth]{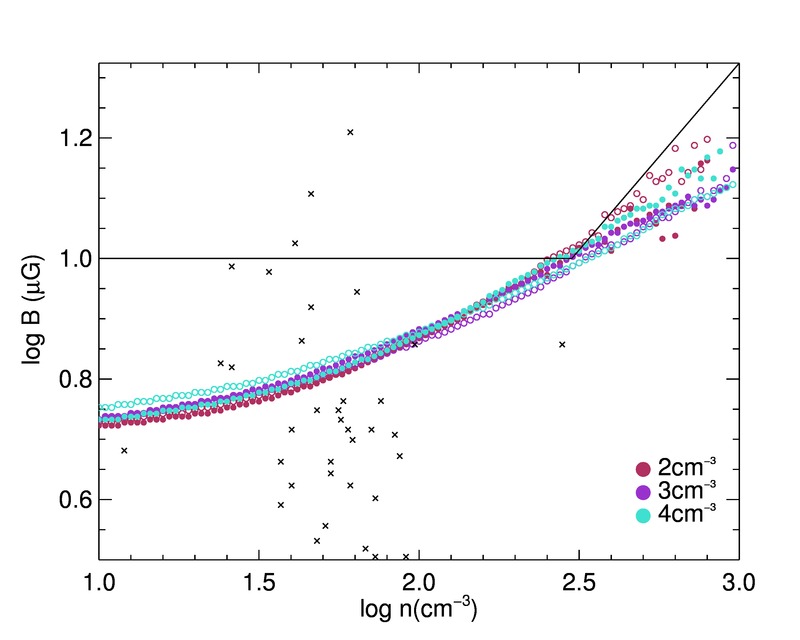}
 \caption{$B$-$n_H$ relation for forced models with $n_0=2$~cm$^{-3}$ as traced by the median of the 2d histogram. Filled(empty) circles are for models without(with) self-gravity. For each initial magnetic field continuous(dashed) lines are for $B$ values containing 25$\%$ of points below and above the median for models without(with) self-gravity. Black lines are the most probable maximum values found by \protect\citet{Crutcheretal10}, while black crosses are the HI data from \protect\cite{HeilesTroland2004}.}
 \label{fig:BroHD}
\end{figure}

Figure \ref{fig:BroHD} shows the $B$-$n$ relation for forced high-density models compared to models with $n_0=2$~cm$^{-3}$. For these figures we used the same time interval as for the figures presented in section \ref{sec:resMain}. As expected, the effects of self-gravity are noticeable at high densities ($n\gtrsim 100$~cm$^{-3}$) where, unexpectedly, for the more massive model the magnetic field intensity grows slower with density, compared with the non self-gravitating case. In fact the slopes of a linear least-square fit for the same density range as the one used in section \ref{sec:resMain} are $\alpha=0.28$, 0.26, and 0.30
for $n_0=2$, 3, and 4~cm$^{-3}$, respectively for the non self-gravitating cases;  while the corresponding values for the self-gravitating models are $\alpha=0.32$, 0.28, and 0.24. Thus, as in section \ref{sec:resMain}, when the relative relevance of self-gravity increases, $B_m$ presents a slower growth with $n$. Note that for the models with increased mass, the behavior of $M_s$ and $M_A$ does not considerably differ from the one we have shown in figures \ref{fig:Mmain} and \ref{fig:MmainG} for $B_0=4.2$~$\mu$G.
\subsubsection{Decaying Turbulence}\label{sec:resHDNF}
\begin{figure}
 \includegraphics[width=\columnwidth]{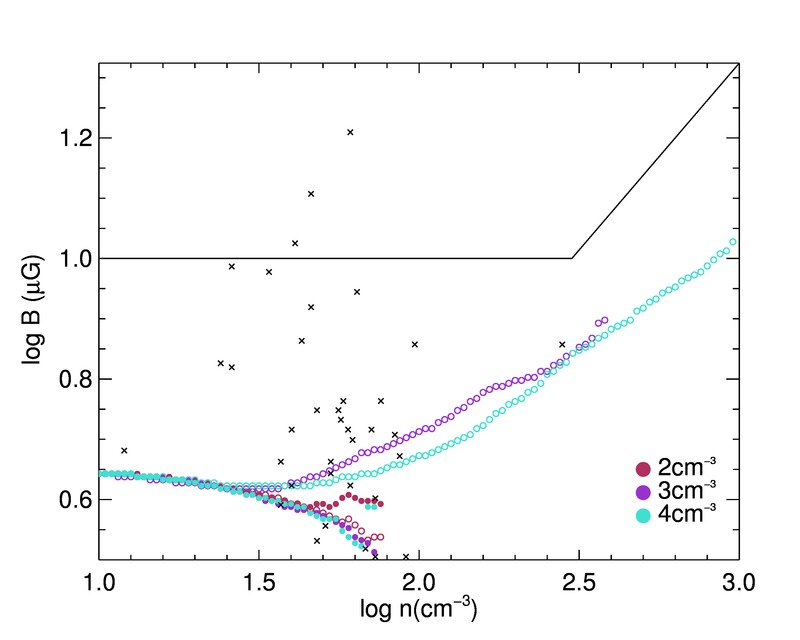}
 \caption{$B$-$n_H$ relation for decaying models with $n_0=2$~cm$^{-3}$, $n_0=3$~cm$^{-3}$, and $n_0=4$~cm$^{-3}$ with (empty circles) and without (filled circles) self-gravity.}
 \label{fig:BroHDNF}
\end{figure}

In the decaying case, self-gravitating massive models develop a $B$-$n$ relation qualitatively different from the one resulting from the $n_0=2$~cm$^{-3}$ simulation. This can be observed in figure \ref{fig:BroHDNF}, where a very slight decay in $B_m$ can be seen for all the models at densities below $\approx 30$~cm$^{-3}$. For non-self gravitating runs and for the self-gravitating standard density model, this decrement continues at higher densities (note that this decay is similar to the one in figures \ref{fig:BromainNF} and \ref{fig:BromainNFG} for the corresponding $B_0$, but it looks stronger due to the different scales we use), but for massive self-gravitating models $B_m$ starts to grow with density at $\approx 30$~cm$^{-3}$. This increase remains for more than an order of magnitude in density.

\begin{figure}
 \includegraphics[width=\columnwidth]{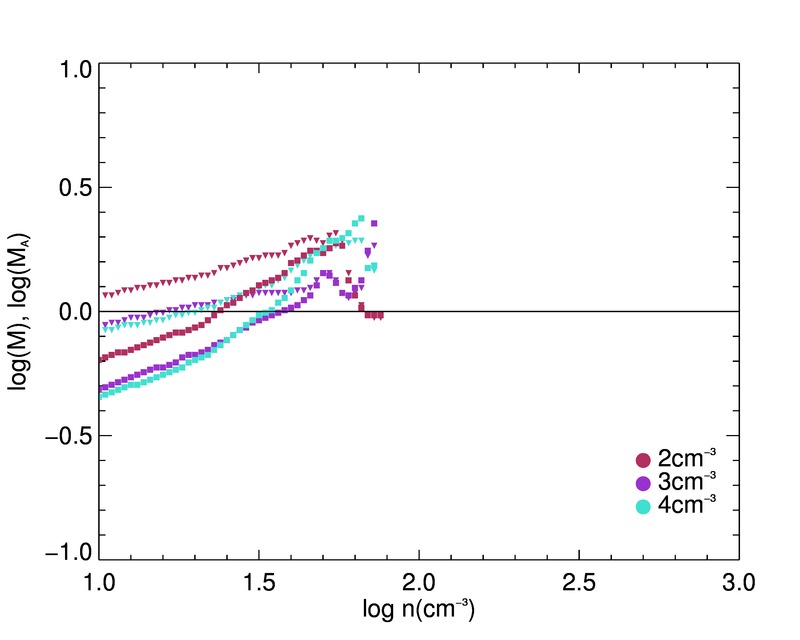}
 \caption{Sonic (triangles) and Alfv\'enic (squares) Mach numbers for decaying models with different $n_0$ without self-gravity.}
 \label{fig:MHDNF}
\end{figure}

\begin{figure}
 \includegraphics[width=\columnwidth]{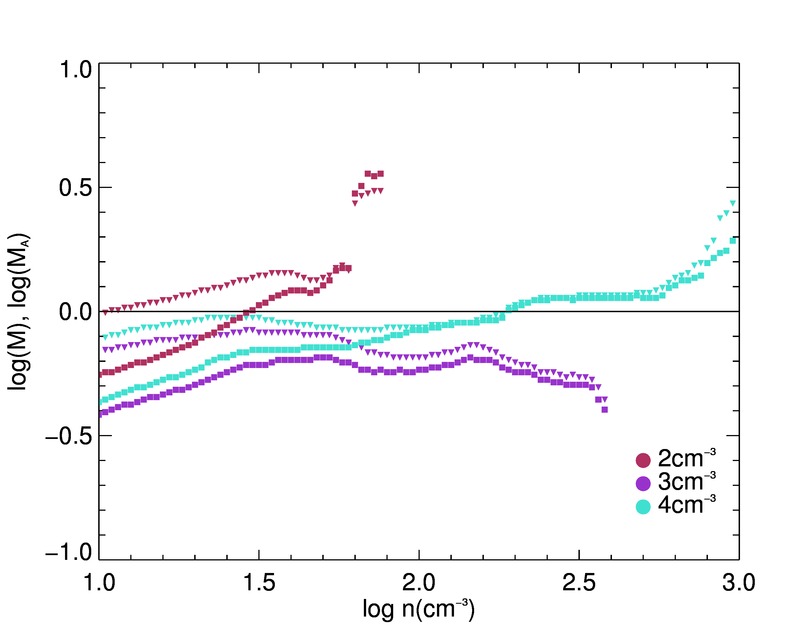}
 \caption{Sonic (triangles) and Alfv\'enic (squares) Mach numbers for decaying self-gravitating models with different $n_0$.}
 \label{fig:MHDNFG}
\end{figure}

From figures \ref{fig:MHDNFG} and \ref{fig:MHDNF}, where the median values of the sonic and the Alfv\'enic Mach numbers  for each density bin resulting from self-gravitating models are plotted, it can be seen that in all the models in this set $M_A\lesssim M_s$, implying that the magnetic pressure is dominant over or balances the thermal pressure. 
Also, as in the models discussed in section \ref{sec:resMainNF}, the simulations for which self-gravity leads to a regime where $B_m$ increases with $n$ are those with predominately subsonic motions at CNM densities. For the most massive case, it also can be seen that: i)At $n\gtrsim 100$~cm$^{-3}$ $M_s\approx M_A $, i.e. magnetic and  thermal pressures are approximately equal. ii) at $n\geqslant 200$~cm$^{-3}$ the gas becomes supersonic and super-Alfv\'enic. The slope of the  $B_m-n$ relation in this case (cyan points in figure \ref{fig:BroHDNF}) is $\approx 0.37$ for  $10^2\leq n \leq 10^3$~cm$^{-3}$, which is consistent with $\alpha\approx\gamma/2$. For the $n_0=3$~cm$^{-3}$ model, the high density gas is near the constant $\beta $ regime, but  does not reach it and remains subsonic and sub-Alfv\'enic, suggesting that the effects of self-gravity are not strong enough to produce an important acceleration of the gas.  
\section{Discussion}\label{sec:disc}
The results described in the previous section show that the magnetic field intensity could have a variety of scalings with the density in a gas with thermal properties similar to those of the local CNM, and formed by the development of thermal instability. Despite this diversity in the $B$-$n$ relation, all the models that we presented have $B$ and $n$ values consistent with at  least part of the observed HI data reported by \cite{HeilesTroland2004} and used by \cite{Crutcheretal10}. The low-density region of the most probable maximum total $B$ value found by the latter authors appears in fact as an upper limit to our models. However, in order to further compare them with observations, a detailed analysis of the physical properties in the resulting individual dense structures is needed.
\subsection{Forced Turbulence}
When the gas is forced to velocities leading to sonic Mach numbers consistent with those derived from observations, magnetic field compression produces median $M_A$ values for the dense gas in the trans-Alfv\'enic or in the slightly super Alfv\'enic regime ($M_A<5$), which are also consistent with observationally obtained values. At the minimum CNM density, we get $M_A$ ranging between $\approx$0.8 and $\approx 3$ (see figures \ref{fig:Mmain} and \ref{fig:MmainG}). Thus, at densities corresponding to the CNM all our forced models are moderately supersonic and slightly super Alfv\'enic. Even for this relatively narrow range in the dense gas $M_A$, we obtain different behaviors for the $B$-$n$ relation in gas with densities similar to those of the CNM. 

The combination of a flat low density regime with an approximate power-law at higher densities, hypothesized by \cite{Crutcheretal10} as the most probable maximum $B$, is recovered for the higher $B_0$ models (whose Alfv\'enic Mach numbers are closer to the observed values), while a continuous increase of the median $B$ with density is observed for the lower  $B_0$ models. Thus, at large enough densities, an increase of $B$ with $n$ is observed for all the initial magnetic field values that we considered. 
This trend with $B_0$ has also been obtained by \cite{Moczetal2017}, who followed the collapse of gas with physical properties similar to those of pre-stellar cores, (i.e. self-gravitating, super-critical, isothermal, highly supersonic gas) with initial magnetic fields corresponding to Alfv\'enic Mach numbers of 35, 3.5, 1.2, and  0.35.  These values are in fact comparable to those obtained for our forced models using the values of the initial Alfv\'en speed and the rms velocity (see table 2 in \cite{Marco17}), which are 11.25, 1.12, 1.05, and 0.50. We observe the same trend with and without self-gravity. Then, the emergence of a combined regime and particularly the appearance of a positive correlation, seems to be independent of the presence of self-gravity.
In fact, in our models the density from which $B$ starts to show an important rise is independent of the presence of self-gravity and does not seem to be modified when more mass is available, at least for $B_0=4.2$~$\mu$G. Even for models with $B_0=8.3$~$\mu$G, in which turbulent compression drives the median $B$ to values close to the most probable maximum found by \cite{Crutcheretal10} of 10~$\mu$G, the rise of $B$ with density starts at $n\approx 30$~cm$^{-3}$, which is approximately  one order of magnitude below the value found by these authors for the switch between the constant $B$ regime and the power-law regime. This is consistent with the $B$-$n$ relation presented in previous numerical works studying molecular cloud formation, where the diffuse gas is sufficiently resolved. \cite{InoueInutsuka16} consider the formation of molecular clouds due to converging magnetized HI flows in the absence of self-gravity, finding that the magnetic field intensity increases with $n$ from relatively low density values. Also \cite{Fogerty2016}, who investigate the role of shear and magnetic fields in molecular cloud formation, report $B$-$n$ distributions consistent with a power-law growth starting $\approx10$~cm$^{-3}$. 
In \cite{Wuetal17}, where the collision of two GMC is addressed, a positive correlation can be observed at densities of about 10~cm$^{-3}$. Note however that \cite{Heitschetal09}, who study the effects of magnetic field on the flow-driven formation of molecular clouds in non self-gravitating models, find  a very weak correlation between $B$ and $n$ for different magnetic field strengths and orientations. The discrepancy with our results could be due to the low $v_{rms}$($M_s$) they get for the cold gas, which are $\approx$0.4-0.15 times the ones we obtain in the forced models, which are consistent with CNM observations.

Without self-gravity, the slope of the logarithmic $B$-$n$ relation for $100\lesssim n \lesssim 300$~cm$^{-3}$ increases with $M_A$. This is consistent with high $M_A$ gas being more prone to magnetic field compressions and with numerical results for a turbulent isothermal gas \citep{PadoanNordlund99,Ostrikeretal01,Burkhartetal09,Collinsetal11}. In the presence of self-gravity, however, the slope is less affected by $B_0$ differences, with low $B_0$ models showing a shallower growth of $B$ with $n$. In fact, for the high $M_A$ models, the presence of self-gravity switches $\alpha$ from   values $>\gamma/2$ to values $<\gamma/2$, which implies that the gas switches from a regime where the magnetic pressure grows with density faster than the thermal pressure to a regime with the opposite behavior.  In the presence of self-gravity  the slope is around 0.3, for the four forced models with $n_0=2$~cm$^{-3}$.
A  shallower power-law in the presence of gravity is also observed for the model with the highest mean density and $B_0=4.2$~$\mu$G but in this case, where for the non self-gravitating model $\alpha<\gamma/2$, the change does not cause a qualitative switch. It is worth noting that the models for which self-gravity produces a shallower
$B$-$n$ relation are those for which \cite{Marco17} find the highest level of alignment between $\mathbf{B}$ and $\nabla n$.  The origin of the reduction in the correlation could be the possible presence of density enhancements not associated with dynamic compressions, presumably associated with motions parallel to the local magnetic field, which we expect to be more relevant when magnetic fields are relatively less important with respect to gravity, i.e. for small $B_0$ or large $n_0$. 

Note that the opposite effect of self-gravity, i.e. a steepening of the $B$-$n$
relation, has been reported by \cite{Collinsetal11}, who found that when no self-gravity is included, the magnetic field scales with density with an exponent 0.4, while after the gravity is turned on, the value becomes 0.48. However, our physical setup  (see section \ref{sec:sim}) is different from  theirs. In fact, they follow the evolution  of an isothermal highly super-critical, highly supersonic ($M_s\approx 10$), and super-Alfv\'enic region because the scope of their work was to study the properties of a turbulent collapsing region. Their setup is similar to the one recently used by \cite{Moczetal2017}, who find a decrease of the scaling index for the strongest initial magnetic field (120~$\mu$G), for which $\alpha$ switches from $\sim 2/3$ to $\sim 1/2$.  

\subsection{Decaying Turbulence}
In the decaying case, the CNM becomes sub-Alfv\'enic at the CNM low density limit for all our models, and we get three different behaviors.
i) No significant correlation between $B_m$ an $n$, besides a very slight decrease of $B_m$, and low ($<100$~cm$^{-3}$) maximal densities. This occurs for two types of models. On one side, the high $B_0$ and $n_0=2$~cm$^{-3}$ models both, with and without self-gravity, where the CNM is supersonic by the time we use for the analysis and for all the density values. On the other side, models with  $n_0=3$ and 4~cm$^{-3}$ without self-gravity, for which the dense gas is predominantly subsonic at the CNM  lower transition density but becomes supersonic at larger but still relatively low densities. For the two types of models with this behavior the CNM becomes super-Alfv\'enic at large enough densities, but $M_A$ remains smaller than or comparable with $M_s$, indicating that magnetic pressure dominates over or balances thermal pressure. 
ii) A very slight decrease of $B_m$ followed by a positive correlation is observed
for high density models including self-gravity, for which the CNM remains mainly 
subsonic and sub-Alfv\'enic (with $M_A\lesssim M_s$) for $n\lesssim 200$~cm$^{-3}$, and a trend towards constant $\beta$ as $n$ increases is observed. 
iii) A clear negative correlation for the low density CNM followed by a positive correlation is observed for low $B_0$ models. By the time we analyze these models, the CNM has become subsonic at the low CNM density limit but, in contrast with the models with the two previous behaviors, the CNM has $\beta\geq 1$. 
 Interestingly, the maximum density with a negative correlation seems to depend more on the $B_0$ value than on the presence of self-gravity, suggesting a MHD origin for the switch on the $B$-$n$ relation. In these models, the presence of self-gravity produces a high density $B$ growth which is better defined and expands to larger densities. 

The first behavior is in contrast with the results reported by previous works for isothermal gas, where supersonic motions do always lead to a positive correlation between $B$ and $n$. However, this difference is somehow expected, because in a cooling gas $M_s>1$ results from the combined effects of motions, which can potentially compress the magnetic field lines, and the lowering temperature, which is not accompanied  by a contraction so that it does not affect the magnetic field intensity.

The second behavior does also differ from what has been found in isothermal gases but in this case the difference can be attributed to the presence of self-gravity in a region with a sufficiently large amount of mass. It is interesting, however that the positive correlation does also develop from $n\approx 30$~cm$^{-3}$. 

The interpretation of the $B_m$ decay observed in ii) and iii) is not straightforward. On one hand, accordingly with \cite{PassotVS03}, this negative correlation suggests the domination of the slow magneto-sonic mode, and the transition to a positive correlation could be due to the transition towards a regime dominated by the fast or the  Alfv\'en mode. However, in order to confirm this origin a detailed analysis of the relative contribution of different MHD modes should be performed. On the other hand, the decay of $B_m$ could also be due to fast turbulent reconnection \citep{LazarianVishniac1999,Lazarianetal2012}. In this case, in order to verify this hypothesis, a detailed study of the parameters characterizing the turbulence which is present in the models where we observe the decay (i.e. injection scale, spectrum scaling) is needed. 

Note that the fact that the second and the third behaviors appear at $M_s<1$ makes them less plausible to occur in the real CNM, at least in the Solar neighborhood. Whether or not they can develop in cold gas with subsonic velocity dispersion depends also on the particular thermal and magnetic properties of that gas. For the third behavior, there is the additional requirement of $\beta>1$ which is also inconsistent with observations characterizing the local CNM as a $\beta<1$ gas. It is interesting, however, that the negative correlation that we report appears in an unforced case, where the  effects of the driving mechanism \citep{Yoonetal2016} do not affect.

\section{Conclusions}\label{sec:conc}

From the results presented in this work we can conclude that:

i) At realistic rms velocities and magnetic field strengths, a cooling gas with thermal pressure and temperature similar to those of the local CNM  develop a positive correlation between the magnetic field intensity and the density starting at $n\lesssim 30$~cm$^{-3}$. The emergence of this positive correlation
does not depend on the presence of self-gravity. The presence of this positive correlation in a gas with supersonic and super-Alfv\'enic motions is consistent with previous results from isothermal simulations.   

ii) In this velocity regime a $B$-$n$ relation constituted by the combination of a flat $B$ at low densities followed by a power-law at high densities develops for a strong enough initial magnetic field intensity. This is consistent with the assumption made by \cite{Crutcheretal10} for the most probable maximum magnetic field intensity according to observational data.

iii) The slope of this positive correlation depends on the relative relevance of self-gravity, but the density at which it emerges seems to be independent of it.
For our --highly gravitationally stable-- models, the effect of self-gravity is to
produce a shallower $B$-$n$ relation in models with either, a relatively weak initial magnetic field or a relatively high mean density. This effect suggests that, for this parameter choice, self-gravity induces motions parallel to the field lines.   

iv) For decaying models with CNM reaching subsonic and sub-Alfv\'enic regimes with $\beta\lesssim 1$, a positive correlation develops only in the presence of self-gravity. The slope of this correlation at large densities is consistent with a constant $\beta(n)$ regime. 

v) For decaying models where the low density CNM is subsonic and sub-Alfv\'enic with $\beta> 1$, we observe a narrow density range with a negative correlation followed by a positive one, presumably both originated by a MHD process.

\section*{Acknowledgements}
We acknowledge the support of UNAM-DGAPA, PAPIIT through the grant: IN110316. This work
has made extensive use of the NASA's Astrophysics Data System Abstract Service.



\bibliographystyle{mnras}
\bibliography{references} 


\bsp	
\label{lastpage}
\end{document}